\documentstyle[prl,aps,epsf,twocolumn]{revtex}
\begin{document}
\title{
Photon Counting Statistics for Single Molecule Spectroscopy
}
\author{Eli Barkai, YounJoon Jung, and Robert Silbey\\
Department of Chemistry,
Massachusetts Institute of Technology,
Cambridge, MA 02139.\\
}
\date{\today}
\maketitle

\begin{abstract}

{\bf Abstract} 

 We present a general relation between  Mandel's $Q$
parameter describing variance of number of photons 
emitted from a single molecule 
and a three time correlation function describing a spectral diffusion
process the molecule is undergoing.
For a prototype of such a  stochastic
process an exact solution  is found which exhibits rich types of physical
behaviors. 

\end{abstract}

Pacs: 78.,  42.50.Ar, 33.80.-b, 05.40.-a \\


 Experimental advances have made it possible
to measure the spectral line of a 
single molecule (SM) embedded in a condensed phase \cite{MO}.
The statistical properties of the photons emitted from 
such a SM illuminates the interplay between dynamical
processes in the condensed phase, SM dynamics and light-matter
interaction.
An important process
responsible for the emission pattern of a SM is
spectral diffusion, 
i.e., perturbers such as two level systems 
or excitations
in the environment of the excited SM lead to 
random changes in the transition frequency of the SM
and in this way the SM comes in and out of resonance with the
exciting laser field
\cite{MAM,Zumb,Reilly2,ZK,Xie,Bawendi,Bach,Orrit,Taras1,PRL}.
The fundamental issue addressed in this work
is the photon statistics
of a single molecule undergoing
a spectral diffusion process. 
Photon statistics of an ensemble of molecules 
will clearly
differ from that of a SM due to inhomogeneous broadening
which is absent in SM spectroscopy (SMS) and due to correlation between fluorescence
photons which exists only on the SM level.

 We will model these fluctuations in a semiclassical way,
using the Bloch
equation, in the limit of weak laser field.
The spectral diffusion process is described
using  a Kubo-Anderson sudden jump approach \cite{Kubo}.
The same model leads to a basic understanding of line
shape phenomena, namely of the average number of
counts $\langle n \rangle$ per measurement time $T$.
Variants of this model have been used in the past fifty years
to model ensemble measurements in e.g., NMR \cite{Kubo} and non-linear
spectroscopy \cite{Mukamel}. 
More recently, this approach was applied to model SMS
in glass,  by Geva and Skinner \cite{Geva} and in \cite{PRL}
to describe static
properties of line shapes and 
by Plakhotnik \cite{Taras1} to model
time dependent fluctuations of SMS.

 Here we present a general formula for Mandel's
$Q$ parameter,
$$ Q= {\langle n^2 \rangle - \langle n \rangle^2 \over \langle n \rangle} - 1 $$
where $n$ is the random number of counts. 
We show how $Q$ is related to the underlying 
stochastic events and how
$Q$ yields
new information not contained in the average $\langle n \rangle$.
For a simple Kubo-Anderson spectral diffusion  process
we find an exact solution for $Q$ which gives insight 
on characteristic behaviors of SMS fluctuations.
If $Q=0$ the photon count
is Poissonian while 
our semiclassical results show a super Poissonian behavior $Q>0$.
For short times (e.g., ns), the correlation function
between fluorescence photons emitted by a single molecule
shows anti-bunching, a sub-Poissonian non-classical effect
 \cite{Or,Zu1}. 
Our approach works well for longer times such that 
$T \Gamma \gg 1$ where $\Gamma$ is the radiative decay rate
of the molecule.
 
 The key quantity in the present formulation is a three time
correlation function $\mbox{C}_3\left( \tau_1 , \tau_2, \tau_3\right)$,
which is similar to non-linear response functions
investigated in the context of four wave mixing processes \cite{Mukamel}.
The three time correlation function contains all the microscopic
information relevant for the calculation of $Q$. It appeared previously
in a recent work of Plakhotnik \cite{Taras1} in the context
of intensity--time--frequency--correlation technique. 
In this letter important
time ordering properties of this function are investigated,
and $Q$ is evaluated.
The relation between $\mbox{C}_3\left( \tau_1 , \tau_2, \tau_3\right)$
and line shape fluctuations described by $Q$ generalizes the Wiener-Khintchine
theorem, which gives the relation between the averaged line shape 
and the one time dipole correlation function. 

 An important issue is the fast and slow modulation limits, to be defined
later.
If the bath is fast the line exhibits the well 
known behavior of motional narrowing (as the time scale of the bath
becomes short; the line is narrowed).
Considering a simple spectral diffusion process,
we show that $Q$ also exhibits motional narrowing,
though in this limit the fluctuations described by
 $Q$ become small and when time scale of the bath
goes to zero we find Poissonian statistics.
If the bath is slow, $Q$ becomes large, and a simple adiabatic type of
approximation, based on steady state solution
of time independent Bloch equation, holds.
Our exact results are evaluated for an arbitrary 
time scale of the bath and are shown to interpolate between the
fast and slow modulation limits.
We note that several authors have analyzed intensity fluctuations
in SMS using the slow modulation assumption \cite{Zumb,Reilly2,ZK,Taras1},
here we extend these results to include
the physically important 
case of fast modulation.

 We assume a simple non-degenerate 
two state single molecule in an external classical laser field. 
The first electronic excited state $|e \rangle$ 
is located at energy $\omega_0$
above the ground state $| g \rangle$.
 We consider the SM Hamiltonian 
\begin{equation}
H = { \hbar \over 2} \omega_0 \sigma_z + \sum_{j=1}^J {\hbar \over 2} \Delta \omega_j (t) \sigma_z -{\bf d}\cdot {\bf E} \cos\left(\omega_{\rm L} t \right),
\label{eq01}
\end{equation}
where $\sigma_z$ is the Pauli matrix. The second term in $H$
reflects the effect of the time evolution of the environment on the absorption
frequency of the SM coupled to $J$ perturbers. The last term in Eq.
(\ref{eq01}) describes the interaction between the SM and the laser field
(frequency $\omega_{\rm L}$), while
%
$ {\bf d} \equiv {\bf d}_{\rm eg} \sigma_x$
is the dipole operator with the real matrix element ${\bf d}_{\rm eg} = 
\langle e | {\bf d} | g \rangle$.

 The molecule is described by $2\times 2$ density matrix $\rho$ whose 
elements are $\rho_{{\rm ee}},\rho_{{\rm gg}},\rho_{\rm eg},\rho_{\rm ge}$. 
Let, 
$u=(\rho_{\rm ge} e^{ - i \omega_{\rm L} t} + \rho_{\rm eg}e^{ i \omega_{\rm L} t})/2$,
$ v = -i( \rho_{\rm ge}e^{ - i \omega_{\rm L} t} - \rho_{\rm eg}e^{i \omega_{\rm L} t})/ 2$,
and
$w=  ( \rho_{\rm ee} - \rho_{gg})/2$.
Using Eq.
(\ref{eq01}) the 
stochastic Bloch equations, in the rotating wave approximation, are 
\begin{eqnarray}
&&  \dot{u} = \delta_{\rm L}(t) v -  \Gamma u/ 2 , \nonumber \\
&& \dot{v} = -\delta_{\rm L}(t) u - \Omega w -   \Gamma v / 2, \nonumber \\
&& \dot{w} = \Omega v - \Gamma w - \Gamma/  2,
\label{eq03}
\end{eqnarray}
$\Omega=- {\bf d}_{\rm eg}\cdot {\bf E}/\hbar$ 
is the Rabi frequency,
$\delta_{\rm L}(t)= \omega_{\rm L} - \omega_0 - \sum_{j=1}^J\Delta \omega_j(t)$
is the detuning frequency
and without loss of generality we set $\omega_0=0$.
SM experiments 
\cite{Lounis,Brunel}  were shown to be compatible with
the {\em deterministic} two level optical Bloch equation approach
and this gives further justification for our assumptions.

According to semiclassical theory of photon counting
statistics \cite{Mandel},
the probability of recording $n$ photons
in time interval $(0,T)$ is given by 
\begin{equation}
 p(n,T) =\left\langle  { W^n \over n!}  \exp\left( - W \right)\right\rangle 
\label{eqMMAA}
\end{equation}
with
$W = \xi \int_0^T I(t) {\rm d} t $,
where $I(t)$ is the photon current and 
$\xi$ is a suitable constant depending on detection
efficiency.  
The semiclassical approach is valid  for large number of
counts, when the stream of incoming absorption
photons (i.e., $\Omega v$) is equal to the stream of emitted fluorescence 
photons  (i.e., $\Gamma \rho_{\rm ee}$).
We use $W\equiv \xi \Omega \int_0^T v(t) {\rm d} t$, which is
$\xi$ times  the work of the driving field  per unit energy defined by 
$\hbar \omega_{\rm L}$, using the Bloch equation it is easy to show
that when $W >>1$ (i.e., photon count is large)
$W \to \xi \Gamma \int_0^T \rho_{\rm ee}(t) {\rm d} t$ as expected.
 Using Eq. (\ref{eqMMAA}) the average number of photons counted in time interval
$(0,T)$ is  $\langle n \rangle=\langle W \rangle$
where $\langle \cdots \rangle$ is an
average over the stochastic process.
Mandel's Q parameter is used to characterize
the fluctuations
and it is straightforward to show that
%
\begin{equation}
Q = { \langle W^2  \rangle - \langle W \rangle^2 \over \langle W  \rangle }.
\label{eqQsol}
\end{equation}
We see that $Q\ge0$ indicating that photon statistics is
super Poissonian.

  We now consider the important
limit of weak laser intensity of  Eq. (\ref{eq03}).
We use a straightforward perturbation
expansion in the Rabi frequency to find
\begin{equation}
v= { \Omega \over 2} \mbox{Re}\left\{  \int_0^t {\rm d} t_1
\exp\left[ - i \int_{t_1}^t {\rm d} t' \delta_{\rm L} (t')  - \Gamma {\left(t - t_1\right)\over 2}
 \right]
\right\}.
\label{eqA03bmmm}
\end{equation}
In standard line shape theories  Eq.
(\ref{eqA03bmmm}) is averaged, the well known result
gives the Wiener--Khintchine formula for the line shape
$\langle I(\omega_{\rm L})\rangle=\lim_{T \to \infty} \langle  W\rangle /T$,
which is valid for stationary processes
\begin{equation}
 \langle   I\left(\omega_{\rm L} \right)  \rangle  = 
{\xi \Omega^2 \over 2} \mbox{Re} \left[ \int_0^{\infty} {\rm d} \tau e^{ - i \omega_{\rm L}
\tau - \Gamma \tau/2} \mbox{C}_1 \left( \tau \right) \right],
\label{eqApzz3mmm}
\end{equation}
where the one time dipole correlation function is
$\mbox{C}_1(\tau) =
\langle e^{ i \int_0^{\tau} \sum_j \Delta_j \omega (t') {\rm d} t ' } \rangle$.
Let us now consider the fluctuation,
$$ \langle W^2 \rangle  = {\xi^2 \Omega^4 \over 16 }  
\int_0^T\int_0^T\int_0^T\int_0^T 
{\rm d} t_1 {\rm d} t_2 {\rm d} t_3 {\rm d} t_4 \times $$
\begin{equation}
 e^{ - i \omega_{\rm L} ( t_2 - t_1 + t_3 - t_4) - \Gamma(|t_1 -t_2| + |t_3 - t_4|)/
2} 
 \mbox{C}_3 \left(\tau_1 , \tau_2 , \tau_3 \right)
\label{eqApzz4}
\end{equation}
where the three time correlation function
 $$ \mbox{C}_3 \left(\tau_1 , \tau_2 , \tau_3 \right)= $$
\begin{equation}
\langle \exp\left[ i \int_{t_1}^{t_2} \sum_j \Delta \omega_j (t') {\rm d} t' -
 i \int_{t_3}^{t_4}\sum_j  \Delta \omega_j (t') {\rm d} t' \right] \rangle,
\label{eqCor3}
\end{equation}
contains the information
on the {\em stationary} spectral diffusion process relevant for the calculation of $Q$.
It depends on the time
ordering of $t_1,t_2,t_3,t_4$ 
for which there are $4!=24$ options.
In Eq. (\ref{eqCor3}) $\tau_1=t_{\rm II}-t_{\rm I}$,
$\tau_2=t_{\rm III}-t_{\rm II}$
and $\tau_3=t_{\rm IV}-t_{\rm III}$ and by definition
$t_{\rm I} < t_{\rm II} < t_{\rm III} < t_{\rm IV}$.
Let us consider the case
 $t_1<t_2<t_3<t_4$ (for this case $t_1=t_{\rm I},t_2=t_{\rm II}$ etc) 
and $\tau_1=t_2 - t_1$, $\tau_2  = t_3 -t_2$ and $\tau_3=t_4 - t_3$.
We define the pulse function
%
%
$S(t) =
-1$ for $t_{\rm I} <  t < t_{\rm II}$, $S(t)=0$
for $t_{\rm II}< t < t_{\rm III}$  and
$S(t)=1$ when $t_{\rm III}< t < t_{\rm I V}$, 
the shape of this pulse is shown in the first line
of Table 1.
Using this pulse the correlation function is written
as a characteristic functional,
$ \mbox{C}_3 \left(\tau_1 , \tau_2 , \tau_3 \right)=
\langle\exp\left[ - i \int_{t_{\rm I}}^{t_{\rm I V}}  S(t') \sum \Delta\omega_j (t') {\rm d} t' \right] 
\rangle.$
Similarly other pulse shapes describe the other time orderings
as described in Table $1$.
We note that  pulses $1$ and $2$  are 
encountered in theory of four wave mixing \cite{Mukamel}.

Eqs. (\ref{eqQsol}-\ref{eqCor3}) 
give a general prescription for the calculation
of $Q$ for broad types of spectral diffusions. 
Now  we investigate basic properties of
line shape fluctuations considering
a simple process \cite{Kubo}.
We assume $J=1$ and $\Delta \omega(t)=\nu h(t)$
and $h(t)$ is a random
telegraph process $h(t)=1$ or $h(t)=-1$. The transition rate from
state up to down and vice versa is $R$, 
generalizations for more complicated cases will be considered elsewhere.

 Using a method of Su\'arez and Silbey \cite{Suarez}, developed  in the
context of photon echo experiments,  we now analyze the
properties of the three time correlation function.
Define the weights
$P^s_{if}(t) = \langle \exp \left[ - i s \int_0^t \Delta \omega (t') {\rm d} t' 
\right] \rangle_{if}$
where the initial 
(final) state of the stochastic process $\Delta \omega$ is $i$
 ($f$)
and $s=0$ or $s=\pm 1$ or $s=\pm 2$ \cite{Weights}.
For example $P^{ - 1}_{++} (t)$
is the value of
$\langle \exp\left( i \int_0^t \Delta \omega (t') {\rm d} t' \right) \rangle$
for a path restricted to have $\Delta \omega (0) =\nu$ and
$\Delta \omega (t) =\nu$.
Using these weights and based on the Markovian property
of the process, we find for $t_1<t_2<t_3<t_4$
$$ \mbox{C}_3 \left(\tau_1 , \tau_2 , \tau_3 \right)= 
{1\over 2} \sum_{i,j,k,l} P^{-1}_{ij}\left( \tau_1 \right)
P^0_{jk}\left( \tau_2 \right) P_{kl}^{1}\left( \tau_3\right) $$
where the summations are over all possible values of $i=\pm$, $j=\pm$ ,$k=\pm$,
and $l=\pm$.

\setlength{\unitlength}{0.35cm}

\vspace{0.5cm}
\begin{tabular}{|c|c|c|c|}
\hline
 $ $ & $S(t)$  &  $\mbox{C}_3(\tau_1,\tau_2,\tau_3)$  & time order\\
\hline
 $ $ &$ $ &$ $ & $ $   \\
1 &
\setlength{\unitlength}{0.35cm}
\begin{picture}(4,3)
\put(0.0,0){-1}
\put(0.2,1){0}
\put(0.2,2){1}
\put(1,0){\line(1,0){1}}
\put(2,0){\line(0,1){1}}
\put(2,1){\line(1,0){1}}
\put(3,1){\line(0,1){1}}
\put(3,2){\line(1,0){1}}
\end{picture} &
${1\over 2}\sum P_{ij}^{-1}(\tau_1)P_{jk}^0(\tau_2) P_{kl}^1(\tau_3)  $ &
$1234_{1\to 4,2\to 3}$ \\
$ $ &$ $ & $ $ & $ $ \\
$2$ &
\setlength{\unitlength}{0.35cm}
\begin{picture}(4,2)
\put(0.2,0){1}
\put(0.2,-1){0}
\put(1,0){\line(1,0){1}}
\put(2,0){\line(0,-1){1}}
\put(2,-1){\line(1,0){1}}
\put(3,-1){\line(0,1){1}}
\put(3,0){\line(1,0){1}}
\end{picture} &
 ${1\over 2}\sum P_{ij}^{1}(\tau_1)P_{jk}^0(\tau_2) P_{kl}^{1}(\tau_3)  $ 
&
$2134_{1\to 4,2\to 3}$ \\
$ $ &$ $ & $ $ & $ $ \\
$3$ &
\setlength{\unitlength}{0.35cm}
\begin{picture}(4,2)
\put(0.2,0){1}
\put(0.2,1){2}
\put(1,0){\line(1,0){1}}
\put(2,0){\line(0,1){1}}
\put(2,1){\line(1,0){1}}
\put(3,1){\line(0,-1){1}}
\put(3,0){\line(1,0){1}}
\end{picture} &
 ${1\over 2}\sum P_{ij}^{1}(\tau_1)P_{jk}^2(\tau_2) P_{kl}^{1}(\tau_3)  $ 
&
$2314_{1\to 4,2\to 3}$ \\
$ $ &$ $ & $ $ & $ $  \\
\hline
\end{tabular}

\vspace{0.2cm}
Table 1.
Each pulse in column $1$ represents four time orderings
corresponding to a three time correlation function
in column $2$.
The symbol $1234_{1 \to 4, 2 \to 3}$ 
corresponds to 
$t_1<t_2<t_3<t_4$,
$t_4<t_2<t_3<t_1$, 
$t_1<t_3<t_2<t_4$ and 
$t_4<t_3<t_2<t_1$. 
The complex conjugate $(\mbox{C.C})$ 
of $\mbox{C}_3\left( \tau_1,\tau_2,\tau_3\right)$
in line $1$ describes the time orderings
$2143_{1 \to 4, 2 \to 3}$,
$\mbox{C.C}\left[ \mbox{C}_3\left( \tau_1,\tau_2,\tau_3\right)\right]$
  of pulse $2$ corresponds to
$1243_{1\to 4,2 \to 3}$, while
$\mbox{C.C}\left[\mbox{C}_3\left( \tau_1,\tau_2,\tau_3\right)\right]$ 
of pulse $3$ 
 corresponds to
$1423_{1 \to 4, 2 \to 3}$.
\vspace{0.5cm}

In a similar way we consider the other $23$ time orderings. We find
only three types of pulses $S(t)$ are needed for the definition
of the three time correlation function, these pulses are
presented in Table 1 together with  the corresponding
$\mbox{C}_3(\tau_1,\tau_2,\tau_3)$.
We then sum all the $24$
contributions  and use the convolution theorem of Laplace transform
to arrive at the main result of this paper
\begin{eqnarray}
 && \langle W^2 \rangle   =
 {\cal L}^{-1} \left\{ \right.  {\Omega^4 \xi^2 \over 16 u^2} \sum_{i,j,k,l} \left[
 \right. \nonumber
 \\
&&  \hat{P}_{ij}^{-1} \left( u + u_{+}\right) \hat{P}_{jk}^0\left(
 u \right) \hat{P}_{kl}^{+1} \left( u +u_{-}\right) \nonumber
\\
 && +  \hat{P}_{ij}^{-1} \left( u + u_{+}\right) \hat{P}_{jk}^0
\left( u + \Gamma \right) \hat{P}_{kl}^{+1} 
\left( u + u_{-}\right)\nonumber \\
&& + \hat{P}_{ij}^{+1} \left( u + u_{-}\right) \hat{P}_{jk}^0
\left( u \right) \hat{P}_{kl}^{+1} 
\left( u + u_{-}\right) \nonumber \\
 && +  \hat{P}_{ij}^{+1} \left( u + u_{-}\right)
 \hat{P}_{jk}^0\left( u + \Gamma \right) \hat{P}_{kl}^{+1} 
\left(  u + u_{-} \right) \nonumber \\
 &&  + 2  \hat{P}_{ij}^{+1} \left( u + u_{-}\right) 
\hat{P}_{jk}^{+2} \left( u + 2 u_{-} \right) 
\hat{P}_{kl}^{+1} \left( u + u_{-} \right) +
 C.C \left. \right]  \left.  \right\}  \nonumber \\
 && \
\label{eqmain}
\end{eqnarray}
where $u_{\pm}=\Gamma/2 \pm i \omega_{\rm L}$,
 ${\cal L}^{-1}$ denotes the inverse Laplace transform
and $\hat{P}^s_{ij}(u)$ is the Laplace transform of
$P^s_{ij}(t)$. 
Using standard methods of complex analysis 
(one can also use standard numerical or symbolic packages) we find the
exact analytical expression for
Mandel's $Q$ parameter
Eq. (\ref{eqQsol}),
based on
Eqs. (\ref{eqmain}) and $\langle W \rangle$ [i.e. Eq. 
(\ref{eqApzz3mmm}) when $T\to \infty$].
It turns out
that $Q$ is not a simple function of the model parameters;
however, as we show below in certain limits simple behaviors
are found.

 We  now
consider the slow modulation limit $\nu \gg \Gamma\gg R$. 
Using our exact result Eq. (\ref{eqmain}) we find
\begin{equation}
Q \sim {\xi \over 2 R} \left( 1 + { e^{- 2  R T} - 1 \over 2 R T}  \right)   { \left( I_{+} - I_{-} \right)^2 \over \left( I_{+} + I_{-} \right) },
\label{eqEXT} 
\end{equation}
where
$I_{\pm} =\Omega^2\Gamma |{\Gamma\over 2} + i( \omega_{\rm L} \mp \nu)|^{-2}/4$
are stationary solutions of time independent Bloch equation
with frequency detuning $ ( \omega_{\rm L} \mp \nu)$, respectively.
Similar to the line shape 
$\langle I \left( \omega_L \right) \rangle \sim (I_{+}+I_{-})/2$, 
$Q$ 
Eq. (\ref{eqEXT}) exhibits splitting with
two peaks at $\omega_{\rm L}=\pm \nu$. 
Unlike the line shape, $Q$ depends on measurement time $T$ and on $R$.
As an example we choose  the parameters 
$\Gamma=40\mbox{MHz}$,
$\Omega=\Gamma/10$, $R=1/\mbox{sec}$
  and
$\omega_{\rm L}=\nu=1\mbox{GHz}$, which mimics 
a SM coupled to a slow two level system in glass
\cite{Orrit}, 
we find for $R T \gg 1$, $Q = 2\times 10^5 \xi$ which 
implies that fluctuations are very large (if compared with the fast modulation
limit soon to be considered).
The lengthly though straightforward
 derivation of Eq. (\ref{eqEXT})
 is 
 based on careful analysis of the poles of $\langle W^2 \rangle$
and $\langle W \rangle$ in Laplace domain.
 The result Eq. (\ref{eqEXT}) can  
be easily  understood in the following way. 
The molecule can be found in two states up $+$ and
down $-$, in the slow modulation limit 
the rate of photon emission in these two states
is determined by the stationary solution of time independent Bloch
equation, namely $I_{\pm}$, while transients are neglected.
Using this physical picture one can re-derive Eq. (\ref{eqEXT}) in a 
straightforward way. 
Eq. (\ref{eqEXT}) describes well the experiments from Bordeaux \cite{Zumb}
of SMS in glass, there the second order
correlation function $g^{(2)}(t)$ was used 
to characterize the dynamics of a SM
coupled to a two level system in the glass. 
To see the relation between these experiments
and our work we recall the relation  \cite{Zu1,Mandel}
$$ Q = \langle I\left( \omega_L \right) \rangle
\left[ { 2 \over T} \int_0^T {\rm d} t_1 \int_0^{t_1} {\rm d} t_2 g^{(2)}( t_2 ) - T \right], $$ 
and for slow process 
$g^{(2)}(t) = 1 + C \exp( - 2 R t )$ 
with $C= (I_{+} - I_{-})^2/(I_{+} + I_{-})^2$,
this expression being equivalent to Eq. (\ref{eqEXT}). The question now
remains whether one can go beyond this slow modulation limit
and investigate faster processes based on SMS.

%

 We therefore consider fast modulation limit for
which only the $T \to \infty$ limit is relevant.
Mathematically different definitions of a
fast modulation exist depending on the $R \to \infty$ limiting
procedure used. 
First consider the case $R \to \infty$ keeping $\nu$
and $\Gamma$ fixed. This fast limit gives $Q=0$.
This is expected since in this case the bath is so fast
the molecule cannot respond to it (e.g., the line width
of the molecule is $\Gamma$).
A more physically  interesting case is to let  $R\to \infty$ 
but keep the ratio 
$\Gamma_{\rm eff}/\Gamma \equiv \nu^2/ \left( R \Gamma\right)$ 
finite. 
In this  fast modulation limit
the  well known line shape is Lorentzian with a width
$ \Gamma + \Gamma_{\rm eff}$
and as mentioned it exhibits motional narrowing.
Using our exact results we find in the same limit
and  for $\omega_{\rm L}=0$ (i.e. at resonance)
\begin{equation}
Q=\frac{ \xi 4\,\Omega^2 \,\Gamma_{\rm eff}^2}{\Gamma\,{\left( \Gamma + \,\Gamma_{\rm eff} \right) }^2\,\left( \Gamma + 2\,\Gamma_{\rm eff} \right) }.
\label{eqFML02}
\end{equation}
To estimate the magnitude of these fluctuations we note that
the  maximum of $Q$ is found
when $\Gamma_{\rm eff}= \Gamma(1 + \sqrt{5})/2$ and then
$Q \simeq 0.36 \xi \Omega^2/\Gamma^2$.
Even if we take $|\Omega|/\Gamma = 0.1$ and $\xi =5\times 10^{-2}$
as a reasonable estimate for weak
laser field and detection efficiency, 
we find a small value  $Q \simeq 1.8\times 10^{-4}$.
However values of $|Q|\le 5\times 10^{-4}$
were reported in \cite{Zu1,remark}, so measurement
in the fast modulation regime might also be possible.

 The fast modulation limit can be analyzed using a method developed
by Loring and Mukamel \cite{Mukamel}  in the 
context of four wave mixing. Briefly,
in this limit we use a factorization approximation
$\mbox{C}_3 \left( \tau_1,\tau_2,\tau_3\right) \simeq 
\mbox{C}_1 \left( \tau_1\right)  
\mbox{C}_1 \left( \tau_2\right)  
\mbox{C}_1 \left( \tau_3\right) $,
and
$\mbox{C}_1 \left( \tau\right)$ is calculated based on 
standard line shape theory  
(details will be given elsewhere \cite{YJ}). This approximation works
well since correlation between the state of the molecule 
(i.e., $+$ or $-$) during 
one pulse interval [say $S(t) = 1$] with that of 
the following pulse interval [say
$S(t^{'}) = 0$] are unimportant, because the bath is fast. Employing this
approach, or using Eq. (\ref{eqmain}) directly, we find
\begin{equation}
Q \simeq \left\{
\begin{array}{c c}
{ \xi \Omega^2 \Gamma_{\rm eff} \over 2 \Gamma \left( \omega_{\rm L}^2 + \Gamma^2_{\rm eff}/4 \right)}
\ & \Gamma \ll \Gamma_{\rm eff}\\
 \ & \ \\
{ \xi \Omega^2 \Gamma_{\rm eff} \Gamma \omega_{\rm L}^2 \over
 \left( \omega_{\rm L}^2 + \Gamma^2 / 4 \right)^3}  \ & \Gamma_{\rm eff} \ll \Gamma .
\end{array}
\right.
\label{eqqQQ}
\end{equation}
Eq. (\ref{eqqQQ}) exhibits several interesting behaviors
{\bf (i)} when $\Gamma \ll \Gamma_{\rm eff}$ a simple relation between the line shape
and $Q$ holds, $Q = 2 \langle I \left( \omega_{\rm L}\right) \rangle/ \Gamma$,
and both exhibit motional narrowing, 
{\bf (ii)} when $\Gamma_{\rm eff} \ll \Gamma$,  
$Q  \sim 1/ R$,
this behavior showing
that as the bath becomes faster the deviations from Poisson statistics
become smaller (similar behavior is
observed in the slow modulation limit) 
{\bf (iii)} a dip in $Q$
is found when $\Gamma_{\rm eff} \ll \Gamma$
(note that we have neglected $O(\Gamma_{\rm eff}/\Gamma)^2$ terms
which give corrections at $\omega_{\rm L}=0$).

  To conclude, the three time correlation function 
describes line shape fluctuations for
a single molecule undergoing a stationary spectral diffusion
process.
We know that similar
correlation functions yield important information
in non--linear spectroscopy \cite{Mukamel} (i.e., beyond linear spectroscopy),
similarly  ``Q SMS" yields detailed information
on dynamical processes in the condensed phase
(i.e., beyond the standard line shape theories).
Depending on the bath 
time scale, different time orderings of the three time correlation
function
contribute to the line shape fluctuations
and in the fast modulation limit all 
time orderings contribute.
Our exact results for 
a simple Kubo-Anderson process
are applicable to the experimental situation, in the slow
and possibly in the intermediate modulation regime (where $Q$ is large).
Our results in the more challenging fast modulation case give
the theoretical limitations on measurement accuracy
needed to detect $Q$. 

{\bf Acknowledgment}
 This work was supported by the NSF. EB thanks M. Orrit
and G. Zumofen 
for discussions.


\begin{thebibliography}{99}
\bibitem{Weights} The weights are
$P_{if}^0(t)=(1/2)[1+(-1)^{\delta_{if}+1}\exp(-2 R t)],$
$P_{+-}^1(t)=P_{-+}^1(t)= R \exp(- R t)\sin(Y_1 t)/Y_1,$
$P_{\pm \pm}^1(t)=\exp(- R t) \left[ \cos\left(Y_1 t \right)
 \mp i \nu \sin\left( Y_1 t \right) / Y_1 \right]$,
$P_{if}^{-1}(t)= \mbox{C.C}\left[ P_{if}^1\left( t \right) \right]$
and $Y_1=\sqrt{\nu^2 - R^2}$. 
$P^{\pm 2}_{if} ( t) $ is given by 
$P^{\pm 1}_{if}(t)$ when $\nu$ is replaced
with $2 \nu$.
\bibitem{remark}  As mentioned the (room temperature) 
experiments in \cite{Zu1} find sub Poissonian
statistics for short measurement times, it is still unclear if and when such
results can be obtained in the presence of spectral diffusion,
our results indicate that in the presence of a very fast
bath spectral diffusion has no effect on photon statistics.
\end{thebibliography}
\end{document}